\documentclass[12pt]{article}
\usepackage{epsfig}

\def\beq{\begin{equation}}
\def\eeq#1{\label{#1}\end{equation}}
\def\eeqn{\end{equation}}

\def\beqa{\begin{eqnarray}}
\def\eeqa#1{\label{#1}\end{eqnarray}}
\def\eeqan{\end{eqnarray}}

\let\bar=\overbar

\def\Dslash{\not{\hbox{\kern-4pt $D$}}}
\def\dslash{\not{\hbox{\kern-2pt $\del$}}}

\def\msb{{\bar{\ssstyle M \kern -1pt S}}}

\def\Title#1{\begin{center} {\Large {\bf #1} } \end{center}}

\begin{document}

\Title{New physics signals in top physics}

\bigskip\bigskip

\begin{raggedright}  

{\it C\'eline Degrande\index{Degrande, C.}\\
Department of Physics, University of Illinois at Urbana-Champaign\\
1110 W Green Street, Urbana, IL 61801\\
Centre for cosmology, particle physics and phenomenology\\
Chemin du cyclotron, 2, Universit\'e catholique de Louvain, Belgium}
\bigskip\bigskip
\end{raggedright}

\section{Introduction}

Effective Field Theory (EFT) provides a model-independent way to search for new physics if the new degrees of freedom are heavy. The new physics appears then as new interactions between the known particles including modification of the SM vertices. In the Lagrangian, they are written as new operators built from the SM fields and invariant under its symmetries. These operators have dimension higher than four and are suppressed by negative powers of the new physics scale $\Lambda$ to get the required dimension for the Lagrangian. Only the operators with the lower dimension, i.e. dimension-six, can be kept in good approximation since the new physics scale is well above the energies probed by the experiments. Consequently, EFT is valid only below the scale of the new physics. In this region, the unitary bound is never reached and no form factors are needed unlike for anomalous couplings (see \cite{Degrande:2012wf} for a complete discussion of the advantages of EFT compared to anomalous couplings). EFT is also more predictive due to the symmetries. For example, an operator often contains several vertices with different numbers of legs all depending on its coefficient only. The symmetries also make the EFT Lagrangian renormalizable and allow loop computation. 
Despite that the number of dimension-six operators that can be added to the SM Lagrangian is large, only a few contribute to a particular process and they can usually be distinguished using several observables. This will be illustrated in the following for top pair production. The complete discussion and list of references can be found in ref. \cite{Degrande:2010kt}. The only references hereafter are new measurements or computations that have been used to update the results from ref.~\cite{Degrande:2010kt}.

\section{The effective Lagrangian for top pair production}

The observables and the squared matrix elements will be expanded to the same order as the Lagrangian, i.e. $\Lambda^{-2}$. Consequently, the new contributions arise as the interference between the dimension-six operators and the SM, in particular, with the dominant QCD processes. The operators should contain top quarks, light quarks and/or gluons. Five operators give the largest contributions to top pair production, one operator which modify the interactions of the top and the gluon : 
\begin{equation}
 \mathcal{O}_{hg} = \left[ \left( H \bar{Q}_L \right) \sigma^{\mu\nu} T^A t_R \right] G^A_{\mu\nu},\label{chromo}
\end{equation}  
two operators involving both the right-handed top and the light quarks :
\begin{eqnarray}
&&\mathcal{O}_{R\, v}  =  \left[ \bar{t}_R \gamma^\mu T^A t_R \right] \sum_{q=u,d}\left[ \bar{q} \gamma_\mu T^A q \right]\nonumber\\
&&\mathcal{O}_{R\, a}  =  \left[ \bar{t}_R \gamma^\mu T^A  t_R \right] \sum_{q=u,d}\left[ \bar{q} \gamma_{\mu}\gamma_5  T^A q \right]\label{4fermion}
\end{eqnarray}
and two similar four-fermion operators with the left-handed doublet of heavy quarks. Those four-fermion operators have to be the product of two color-octet current to interfere with the SM and can only affect top pair production by quark annihilation. The full Lagrangian is written as
\begin{equation}
{\mathcal L_{t\bar{t}}}  =  {\mathcal L}^{SM}_{t\bar{t}} +  \frac{1}{\Lambda^2} \left( c_{hg} \left({\mathcal O}_{hg} + h.c.\right)+
\left(c_{R\, v} {\mathcal O}_{R\, v} + c_{R\, a} {\mathcal O}_{R\, a} + R\leftrightarrow L\right) \right).
\end{equation}	

\section{Phenomenology}

\begin{figure}[htb]
\begin{center}
\epsfig{file=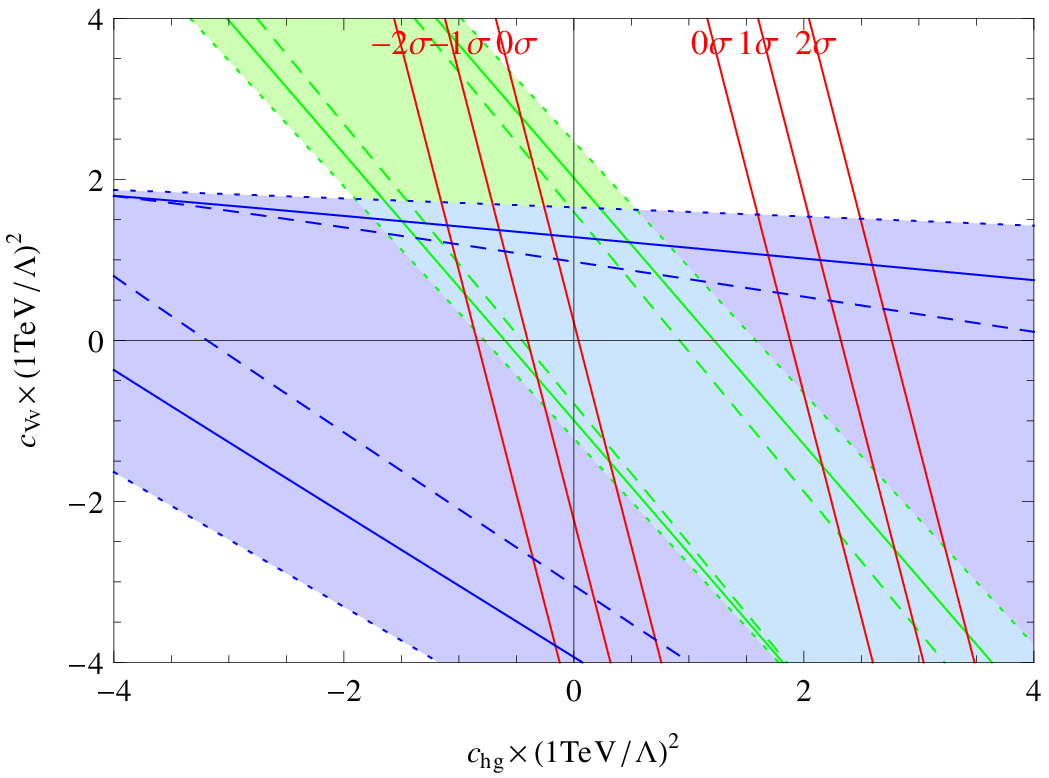,width=0.49\textwidth}
\epsfig{file=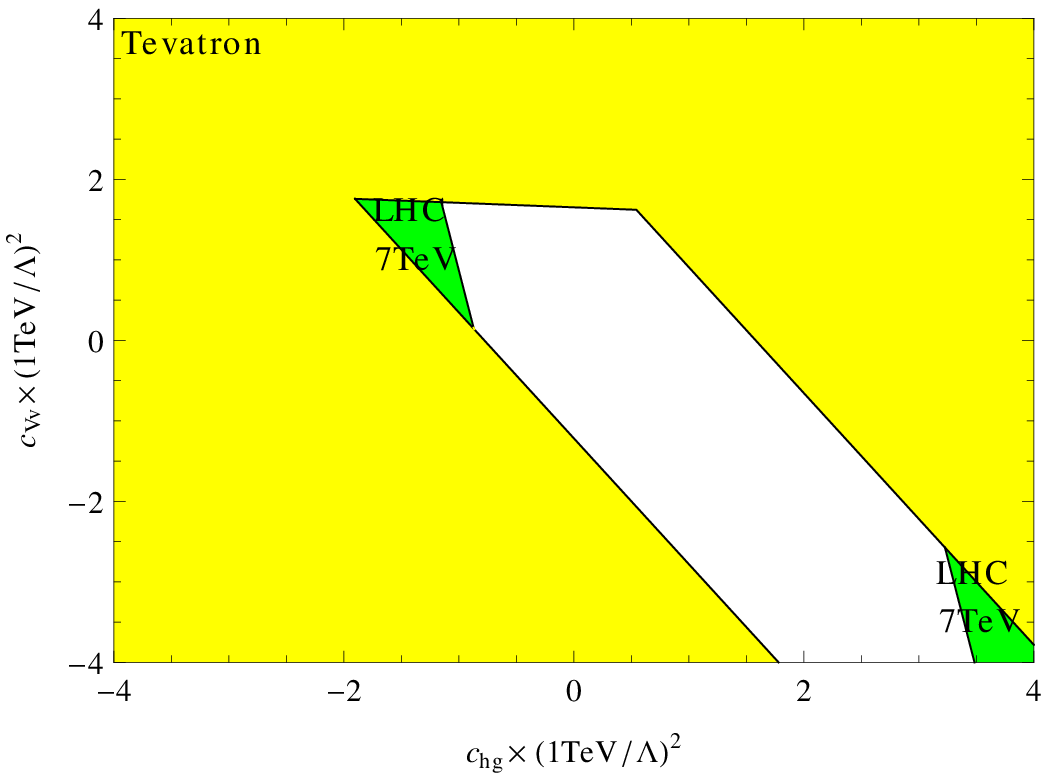,width=0.49\textwidth}
\caption{On the left, the allowed regions from the LHC total cross-section at 7 TeV (red lines) and the Tevatron total cross-section (green region) and invariant mass shape (blue region) in the $c_{Vv}$-$c_{hg}$ plane. On the right, the summary of the present constraints.}
\label{fig:LHCTconst}
\end{center}
\end{figure}

The total cross-section at hadron colliders only depends on two parameters $c_{hg}$ and $c_{Vv}=c_{Rv}+c_{Lv}$. The axial operators $\mathcal{O}_{R\, a}$ and $\mathcal{O}_{L\, a}$ can only contribute to observables that are odd functions of the scattering angle because the axial current is odd under the exchange of the light quark and anti-quark. They have thus no effect on the total cross-section. The contributions from the two vector operators are identical since the cross-section is not sensitive to the helicity of the top. The constraints from the measurements of the total cross-section at the Tevatron and at the LHC \cite{Atlas:2012xs} are shown on Fig.~\ref{fig:LHCTconst}. The new NNLO results for the top pair production by quark annihilation \cite{Baernreuther:2012ws} have allowed to significantly reduce the region allowed by the Tevatron data. The four-fermion operators have a small contribution at the LHC where the dominant production mechanism is gluon fusion. Consequently, the constraints at the LHC in the $c_{Vv}$ direction is less stringent than at the Tevatron.  

%%%%%%%%%%%%%%%%%%%%%%%%%%%%%%%%%%%%%%%%%%%%%%%%%%%%%%%%%%%%%%%%%%%%%%%%%%%
%%%%%%%%%%%%%%%%%%%%%%%%%%%%%%%%%%%%%%%%%%%%%%%%%%%%%%%%%%%%%%%%%%%%%%%%%%%

Dimension-six operators are also expected to influence the shape of the distributions if the $\Lambda^{-2}$ is balanced in the numerator by the energies of the process. The contributions of the four-fermion operators have an extra $s$ factor compared to the SM one. Consequently, they can be further constrained using the the invariant mass distribution. On the contrary, the contribution from the operator ${\mathcal O}_{hg}$ has to be proportional to the Higgs vev and to the top mass and mainly affects the overall normalization. While the shape distortion are too small at the LHC so far, the Tevatron strongly constrain the four-fermion operators contribution as shown on Fig.~\ref{fig:LHCTconst}.

\begin{figure}[htb]
\begin{center}
\epsfig{file=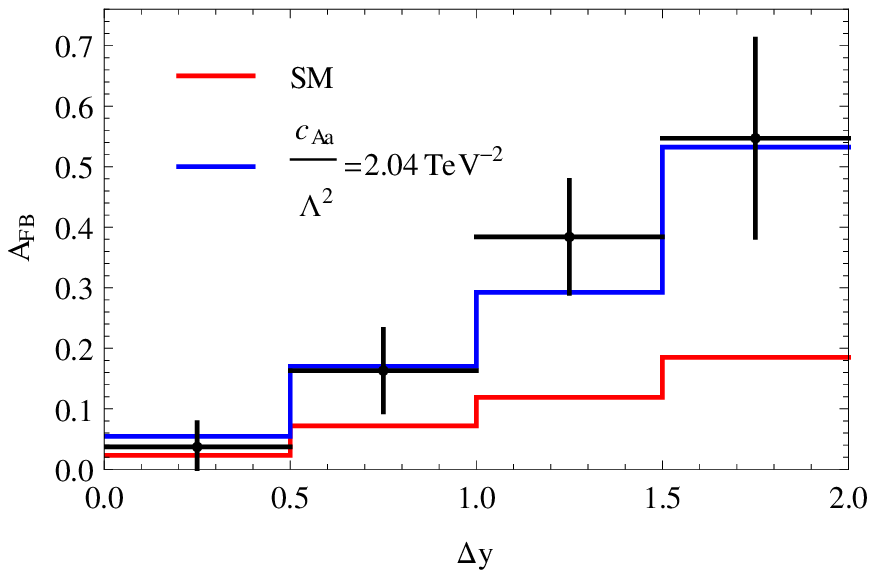,width=0.49\textwidth}
\epsfig{file=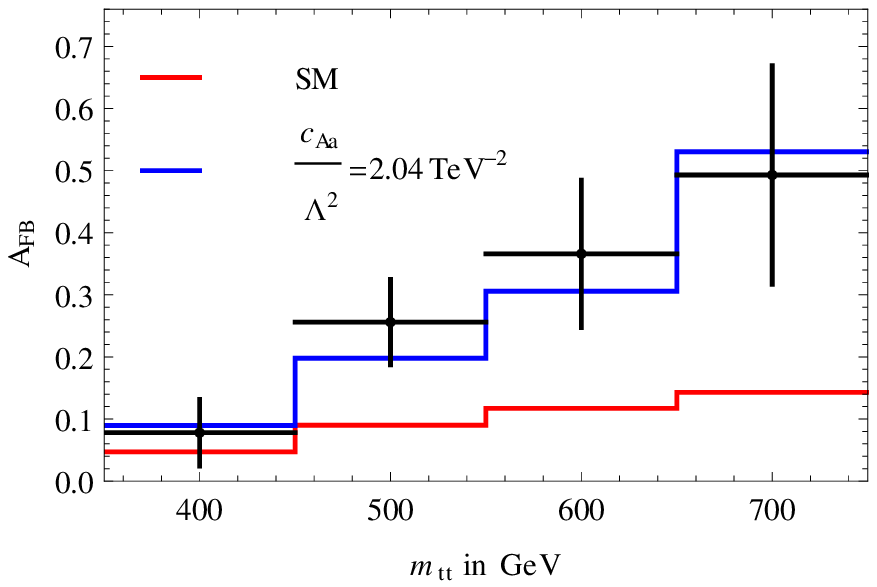,width=0.49\textwidth}
\caption{Distributions of the forward-backward asymmetry as a function of the rapidity (left) and invariant mass (right). The SM predictions are in red, the SM and the interference with the dimension-six operators predictions for the value of $c_{Aa}/\Lambda^2$ which gives the best fit for the total asymmetry at the Tevatron are in blue, the data \cite{CDF:2012afb} are in black.}
\label{fig:Afbdist}
\end{center}
\end{figure}

The forward-backward asymmetry is an odd function of the scattering angle and allows to constrain the axial operators. In fact, all the measurements related to the asymmetry depend only on $c_{Aa}=c_{Ra}-c_{La}$ as long as the spin of the top quarks are not measured simultaneously. The best value obtained by fitting the total asymmetry measured by CDF~\cite{CDF:2012afb}, $c_{Aa}/\Lambda^2=2.04$ TeV$^{-2}$, can then be used to predict the rapidity and invariant mass distributions. As shown on Fig.~\ref{fig:Afbdist}, the predictions are in good agreement with the data. Furthermore, both the forward-backward asymmetry and the invariant mass distribution can be explained as they are affected by independent combinations of operators.

%%%%%%%%%%%%%%%%%%%%%%%%%%%%%%%%%%%%%%%%%%%%%%%%%%%%%%%%%%%%%%%%%%%%%%%%%%%
%%%%%%%%%%%%%%%%%%%%%%%%%%%%%%%%%%%%%%%%%%%%%%%%%%%%%%%%%%%%%%%%%%%%%%%%%%%

Spin dependent observables are required to disentangle the four-fermion operators with the right-handed top from those with the left-handed top. Fortunately, spin correlation can be observed in top pair production. The parameter sensitive to the amount of top pair produced with the same or opposite helicity has been measured~\cite{ATLAS:2012ao} and the precision is close to the expect effects from the dimension-six operators. However, this parameter is only sensitive to  $c_{Vv}$ and $c_{hg}$. The other parameters vanish in the SM but are sensitive to the orthogonal combination $c_{Av}=c_{Rv}-c_{Lv}$. Again, those effects due only to the four-fermion operators are small at the LHC. 

\paragraph{}To sum up, effective field theories provide a model independent way to search for heavy new physics and have many desirable properties. They are also useful for phenomenology at hadron colliders, various dimension-six operators are already constrained by their data. Those constraints can be improved by new measurements and more precise predictions for the SM contributions.

\bigskip
The author is a fellow of the Fonds National de la Recherche Scientifique and the Belgian American Education Foundation.

\def\Discussion{
\setlength{\parskip}{0.3cm}\setlength{\parindent}{0.0cm}
     \bigskip\bigskip      {\Large {\bf Discussion}} \bigskip}
\def\speaker#1{{\bf #1:}\ }
\def\endDiscussion{}

\end{document}